
\input harvmac
\overfullrule=0mm

%
\def\frac#1#2{\scriptstyle{#1 \over #2}}

%
%

%
\def\({ \left( }
\def\){ \right) }
%


\def\IR{\relax{\rm I\kern-.18em R}}
\font\cmss=cmss10 \font\cmsss=cmss10 at 7pt
\def\IZ{\relax\ifmmode\mathchoice
{\hbox{\cmss Z\kern-.4em Z}}{\hbox{\cmss Z\kern-.4em Z}}
{\lower.9pt\hbox{\cmsss Z\kern-.4em Z}}
{\lower1.2pt\hbox{\cmsss Z\kern-.4em Z}}\else{\cmss Z\kern-.4em Z}\fi}
\def\inbar{\,\vrule height1.5ex width.4pt depth0pt}
\def\IB{\relax{\rm I\kern-.18em B}}
\def\IC{\relax\hbox{$\inbar\kern-.3em{\rm C}$}}
\def\ID{\relax{\rm I\kern-.18em D}}
\def\IE{\relax{\rm I\kern-.18em E}}
\def\IF{\relax{\rm I\kern-.18em F}}
\def\IG{\relax\hbox{$\inbar\kern-.3em{\rm G}$}}
\def\IH{\relax{\rm I\kern-.18em H}}
\def\II{\relax{\rm I\kern-.18em I}}
\def\IK{\relax{\rm I\kern-.18em K}}
\def\IL{\relax{\rm I\kern-.18em L}}
\def\IM{\relax{\rm I\kern-.18em M}}
\def\IN{\relax{\rm I\kern-.18em N}}
\def\IO{\relax\hbox{$\inbar\kern-.3em{\rm O}$}}
\def\IP{\relax{\rm I\kern-.18em P}}
\def\IQ{\relax\hbox{$\inbar\kern-.3em{\rm Q}$}}
\def\IGa{\relax\hbox{${\rm I}\kern-.18em\Gamma$}}
\def\IPi{\relax\hbox{${\rm I}\kern-.18em\Pi$}}
\def\ITh{\relax\hbox{$\inbar\kern-.3em\Theta$}}
\def\IOm{\relax\hbox{$\inbar\kern-3.00pt\Omega$}}




\def\Ga{\alpha}\def\Gb{\beta}\def\Gc{\gamma}\def\GC{\Gamma}
\def\Gd{\delta}

\def\Gl{\lambda}

\def\Gs{\sigma}\def\GS{\Sigma}\def\Gt{\theta}


\def\ch{{\rm ch}}


\def\SF{{\underline{\smash{F}}}}
\def\SN{{\underline{\smash{\nu}}}}
\def\SS{{\underline{\smash{S}}}}
\def\SK{{\underline{\smash{K}}}}
\def\encadre#1{\vbox{\hrule\hbox{\vrule\kern8pt\vbox{\kern8pt#1\kern8pt}
\kern8pt\vrule}\hrule}}
\def\encadremath#1{\vbox{\hrule\hbox{\vrule\kern8pt\vbox{\kern8pt
\hbox{$\displaystyle #1$}\kern8pt}
\kern8pt\vrule}\hrule}}

\Title{SPhT/92-162 hep-th/9212108}
{{\vbox {
\bigskip
\centerline{A Generating Function for Fatgraphs} }}}

\bigskip
\centerline{P. Di Francesco}
\bigskip
\centerline{and}
\bigskip
\centerline{C. Itzykson,}

\bigskip

\centerline{ \it Service de Physique Th\'eorique de Saclay
\footnote*{Laboratoire de la Direction des Sciences et
de la Mati\`ere du Commissariat \`a l'Energie Atomique.},}
\centerline{ \it F-91191 Gif sur Yvette Cedex, France}

\vskip .5in
We study a generating function for the sum over fatgraphs with
specified valences of vertices and faces, inversely weighted
by the order of their symmetry group.
A compact expression is found for general
(i.e. non necessarily connected) fatgraphs.
This expression admits a matrix integral
representation which enables to  perform semi--classical
computations, leading in particular to a closed formula
corresponding to (genus zero, connected) trees.

\noindent
\Date{11/92}


\lref\BEL{G.V. Belyi "On Galois extensions of a maximal cyclotomic
field" Math. USSR Izvestija {\bf 14}(1980) 247-256.}
\lref\VS{V.A. Voevodsky and G. Shabat "Drawing curves over number fields"
in "A. Grothendieck Festschrift", Birkh\" auser.}
\lref\JW{ J. Wolfart "Mirror-invariant triangulations of Riemann surfaces,
triangle groups and Grothendieck dessins : variations on a thema of Belyi",
Mathematics Seminar der Universit\" at Frankfurt (1992).}
\lref\BI{ M. Bauer and C. Itzykson "Triangulations" to appear in
Publications de la R.C.P. 25, Strasbourg.}
\lref\PEN{ R.C. Penner "Perturbative series and the moduli space of
Riemann surfaces" J. Diff. Geom. {\bf 27} (1988) 35-53.}
\lref\ZI{C. Itzykson and J.-B. Zuber "Matrix integration and
combinatorics of the modular groups" Comm. Math. Phys. {\bf 134}
(1990) 197-207.}


\noindent{ \bf 1.} The relation between fatgraphs describing finite
cellular decompositions of compact orientable surfaces and arithmetic
curves originates in a theorem of Belyi \BEL\ and is described
in detail in recent reviews \VS \JW \BI.
It raises the difficult but interesting problem of understanding the
action of the Galois group of $\bar {\IQ}$ over $\IQ$ on
fatgraphs.
The fatgraphs arise naturally in the Feynman diagrammatic expansion
of random matrix models.
They consist of ribbons  bordered by parallel
double lines, joining at vertices and defining faces on a compact
orientable surface of
given genus.
The action of the group $Gal({\bar {\IQ}}/\IQ)$ preserves the numbers
$S_v$, $F_v$ of vertices and faces of valency $v$ for each $v \geq 1$,
satisfying
\eqn\const{ \sum v \ S_v \ = \ \sum v \ F_v \ = \ 2A}
with $A$ the number of edges. For convenience of notation we write
$\SS$, $\SF$ for sequences $\{S_v\}$, $\{F_v\}$
satisfying the above relation.
The fatgraphs can be built as in a lego game with basic element
a half edge (there are $2A$ of them). In the following,
we reinterpret the sequences
$\SS$ and $\SF$ as conjugacy classes of the group of permutations
of the $2A$ half edges, $\GS_{2A}$.
Each fat graph (connected or not) admits an automorphism
group $H$ of order $h$ ($h$ is a divisor of $2A$ in the connected case)
as well as a cartographic group (of order a multiple of $2A$ in the
connected case).
The latter is generated by three permutations $\Gs_0$, $\Gs_1$, $\Gs_2$
in $\GS_{2A}$, satisfying
$$ \Gs_2 \Gs_1 \Gs_0 = {\rm id} $$
and belonging respectively to the classes $\SS$, $[2^A]$ and $\SF$.
The element $\Gs_0$ permutes circularly the half edges around each
vertex, $\Gs_1$ interchanges half edges and $\Gs_2$ permutes circularly
half edges around faces.
Up to overall conjugacy in $\GS_{2A}$, the
automorphism group $H$ of a fatgraph is the commutant of
$\{ \Gs_0,\Gs_1,\Gs_2\}$ in $\GS_{2A}$.

Let us look for a generating function for the quantities
\eqn\funcz{ z(\SS,\SF)= \sum_{{\cal G}_{\SS,\SF} }
{ 1 \over h({\cal G}_{\SS,\SF})} }
where the sum on the r.h.s. runs over all fatgraphs with given
assignments $\SS$ and $\SF$ of vertices and faces subject to
the constraints \const, and $h$ is the order of the automorphism group
of the fatgraph.
If the sum is taken only over {\it connected} fatgraphs we define a
similar and more relevant quantity
\eqn\funcf{ f(\SS,\SF)= \sum_{{\cal G}_{\SS,\SF}\ {\rm connected}}
{ 1 \over h({\cal G}_{\SS,\SF})} }

This definition makes clear the following criterion.
Given a connected fatgraph or equivalently a finite
cellular decomposition of an orientable compact connected surface
with specified  $\SS$, $\SF$ and order $h$ of its automorphism group,
the equality
$$ h f(\SS,\SF) =1$$
ensures that the corresponding curve is defined on $\IQ$.
This criterion is sufficient but by no means necessary.
More generally the number of conjugates of the curve
(under ${Gal}({\bar {\IQ}} /{\IQ})$) is bounded by $h f(\SS,\SF)$.

\bigskip

\noindent{ \bf 2.} To compute the quantity $z(\SS,\SF)$ defined in
\funcz,
we consider the following problem: count the number $N_{\SS,\SF}$
of triplets of elements $\Gs_0$, $\Gs_1$, $\Gs_2$ in $\GS_{2A}$,
such that
$$\eqalign{
(i)\ \ & \Gs_2 \Gs_1 \Gs_0 = {\rm id} \cr
(ii)\ \  & \Gs_0 \in \SS, \ \ \Gs_1 \in [2^A], \ \  \Gs_2 \in \SF \cr}$$
This is a particular case of the following situation.
Let $C_i$ be the conjugacy classes in a finite group $\GC$,
$r$ the irreducible representations of $\GC$ over $\IC$
forming the
dual $\widetilde{ \GC}$ of the group,
$\chi_{r}(C)$ the value of the irreducible
character in the representation $r$ over the class $C$ and finally
$dim_r \equiv \chi_r({\rm id})$ the dimension of $r$.
Set
$$
N_{C_2,C_1,C_0}\  = \ \eqalign{&\hbox{number of triplets
$(g_2,g_1,g_0)$ such that} \cr
&\hbox{$g_i \in C_i$ and $ g_2 g_1 g_0 = 1$.} \cr}$$

\bigskip

\noindent{\bf Lemma 1. \ }(Frobenius)
$$\encadremath{ N_{C_2,C_1,C_0} =
{|C_2||C_1||C_0| \over|\GC|} \sum_{r \in {\widetilde {\GC}}}
{{\chi_r(C_2) \chi_r(C_1) \chi_r(C_0) } \over dim_r} }$$

\noindent{\bf Remark:} these non negative integers are directly
related to the structure constants of the (commutative) algebra
of classes
$$ C_2 \ C_1 \ = \sum_{C_0} \ N_{C_2,C_1,{\bar {C_0}}} \ C_0$$
where ${\bar {C_0}}\equiv \{ g \in \GC | g^{-1} \in C_0\}$.

If $|\SS|$, $|[2^A]|$ and $|\SF|$ stand respectively
for the number of elements in the classes $\SS$, $[2^A]$ and
$\SF$, we have
$$ N_{\SS,\SF} \equiv N_{\SS,[2^A],\SF} =
{|\SS||[2^A]||\SF| \over {(2A)!}}
\sum_{r \in {\widetilde {\GS}}_{2A}}
{\chi_r(\SS) \chi_r([2^A]) \chi_r(\SF) \over dim_r} $$
Each fatgraph $\cal G$ characterized by $\SS$, $\SF$ corresponds
to a triplet $\Gs_2$, $\Gs_1$, $\Gs_0$ as above, up to
simultaneous conjugacy in $\GS_{2A}$ and the number of
{\it distinct} conjugates is $(2A)!/h({\cal G})$
as follows from the
definition of the automorphism group $H$ of $\cal G$, of order
$h({\cal G})$.
Therefore we have

\bigskip

\noindent{ \bf Lemma 2.}
$$\encadremath{ \eqalign{
z(\SS,\SF) &= \sum_{{\cal G}_{\SS,\SF}}
{1 \over h({\cal G}_{\SS,\SF}) } = { N_{\SS,\SF} \over {(2A)!}}
\cr
&= {(2A-1)!! \over { \prod_v v^{S_v+F_v} S_v! F_v!}}
\sum_{r \in {\widetilde { \GS}}_{2A}}
{\chi_r(\SS) \chi_r([2^A]) \chi_r(\SF) \over dim_r} \cr} }$$
where we have used the fact that the number of elements in
a class $\SK\equiv[\prod v^{k_v}]$ in $\GS_{2A}$ is equal to
$$ |\SK|= { (2A)! \over {\prod_v v^{k_v} k_v!}}$$
We know of no such explicit formula for the function $f(\SS,\SF)$
of eqn.\funcf, i.e. when the sum is restricted to connected graphs.
Note that $z(\SS,\SF)$ is invariant under the interchange
$\SS \leftrightarrow \SF$.

\bigskip

\noindent{ \bf 3. } To define a generating function for
the quantities $z$ and $f$, we introduce two infinite sequences
of variables
$$ t \equiv \{ t_1,t_2,...\} \ \  t' \equiv
\{ t_1', t_2',... \} $$
For short we write $t^{\SS}$ for the monomials $t_1^{S_1}t_2^{S_2}..$
and similarly for $t'^{\SF}$.
The generating ("partition") function $Z(t,t')$ is the
formal series
$$ Z(t,t')= \sum_{\SS,\SF}\  t^{\SS}\  t'^{\SF}\  z(\SS,\SF)$$
where the sum is restricted by the condition
$\sum v S_v = \sum v F_v \equiv 0\  {\rm mod} \ 2$.
A glance at lemma 2 suggests however a better choice of
variables
$$ \Gt \equiv \{ \Gt_k={t_k \over k}\} \ \ \Gt_k' \equiv
\{ \Gt'={t_k' \over k} \} $$
Without changing the notation, we will consider
$Z$ as a function of $\Gt$, $\Gt'$ and write abusively
$Z(\Gt,\Gt')$.
The representations of $\GS_{2A}$ are indexed by Young tableaux
with $2A$ boxes so that we can trade the index $r$ appearing
in lemma 2 for a Young tableau $Y$.
The same tableau pertains to a representation of the
linear group $GL(N)$ for $N$ large enough and one defines the
corresponding characters as generalized Schur functions through the
Frobenius reciprocity formula
\eqn\frob{ ch_{Y}(\Gt)= \sum_{{\rm classes} \ \SN \in \GS_{|Y|}}
\chi_{Y}(\SN) {\Gt_1^{\nu_1} \over \nu_1!}{\Gt_2^{\nu_2} \over \nu_2!}
\cdots {\Gt_{|Y|}^{\nu_{|Y|}} \over \nu_{|Y|}!} }
This leads to
\eqn\exzz{\eqalign{ Z(\Gt,\Gt')&=\sum_{A \geq 0} Z_A(\Gt,\Gt') \cr
Z_A(\Gt,\Gt')&=\sum_{|Y|=2A} ch_{Y}(\Gt){{(2A-1)!! \chi_Y([2^A])}
\over {\chi_Y([1^{2A}])}} ch_Y(\Gt') \cr}}
with $Z_0=1$, and the convention $(-1)!!=1$. Recall that the Schur
polynomials $p_n(\Gt)$ are defined through
$$ e^{\sum_{i=1}^{\infty} z^i \Gt_i } =
\sum_{n=0}^{\infty} z^n p_n(\Gt) $$
and that the generalized Schur functions for a Young tableau $Y$
with $f_i$ boxes in the $i$th line,
$f_1 \geq f_2 \geq ...\geq f_{2A} \geq 0$, $\sum f_i=2A$, read
\eqn\genschur{ch_Y(\Gt)= \det \big[ p_{j-i+f_i}(\Gt)
\big]_{1 \leq i,j \leq 2A}}
The above expressions enable one to easily compute the quantity under
brackets in $Z_A$. Namely attach to $Y$ the strictly decreasing sequence
of non negative integers
\eqn\lis{ l_i = f_i+2A-i \ \ \ \ \ 1 \leq i \leq 2A}
For lack of a better name
we shall say that a Young tableau is "even" if the number
of odd $l's$ equals the number of even ones in the list \lis.
We have

\bigskip

\noindent{\bf Lemma 3.}

$$\encadremath{
\eqalign{\phi_Y\equiv
{{(2A-1)!! \chi_Y([2^A]) } \over {\chi_Y([1^{2A}])}} &=
(-1)^{{A(A-1)} \over 2} { { \prod_{l \ {\rm odd}}\ l!! \
\prod_{l' \ {\rm even}} (l'-1)!! } \over {\prod_{l \ {\rm odd} \atop
l' \ {\rm even}} (l-l') }} \ \ \hbox{if $Y$ is even}\cr
&= 0 \ \ \ \ \ \hbox{otherwise.} \cr}}$$

\noindent{\bf Remark :} when it is not zero,
the result of lemma 3 is always
a relative odd integer.

\noindent{}The proof of lemma 3 uses the
reciprocity formula \frob\ and the determinant form \genschur\ of the
generalized Schur function.
For instance the denominator $\chi_Y([1^{2A}])$
in the expression computed in lemma 3,
is the coefficient of $\Gt_1^{2A}/(2A)!$ in the expansion
\frob. {}From \genschur\ it has a simple determinant form, which
immediately gives
$$ \chi_Y([1^{2A}]) = {{(2A)!} \over {l_1!l_2!...l_{2A}!}}
\prod_{1 \leq i<j \leq 2A} (l_i - l_j) $$
The numerator of the expression of lemma 3 is proportional to the
coefficient of $\Gt_2^A / A!$ in the expansion \frob. Collecting
both terms yields the desired result.

The reader will easily convince him(her)self that the following property
holds.
Let $Y\equiv \{ f_1\geq f_2 \geq ... \geq f_{2A}\geq 0\}$
be an even Young tableau. Define $l_i$ as in \lis.
Complete if necessary the $f$ sequence with zeroes
or delete some zero $f$'s so that the same sequence
now reads $f_1\geq f_2 \geq ...\geq f_{2N}\geq 0$ for
some $N$. Define now $\{L_i=f_i+2N-i \ , \ 1\leq i \leq 2N\}$
as in \lis, for this new but equivalent sequence (both
pertain to the same tableau $Y$). In particular the
evenness of $Y$ does not depend on $N$, it is an intrinsic
property of the representation.
Moreover we have

\bigskip

\noindent{\bf Lemma 4.}

$$\encadremath{\phi_Y=
(-1)^{{A(A-1)} \over 2} {{\prod_{l \ {\rm odd}} \ l!! \
\prod_{l' {\rm even}} (l'-1)!!} \over {\prod_{l \ {\rm odd} \atop
l' \ {\rm even}} (l-l') }} =
(-1)^{{N(N-1)} \over 2} {{\prod_{L \ {\rm odd}} \ L!! \
\prod_{L' {\rm even}} (L'-1)!!} \over {\prod_{L \ {\rm odd} \atop
L' \ {\rm even}} (L-L') }} }$$

This lemma simplifies calculations. For instance, if $Y$ is a single
row with $2A$ boxes, it is even, we can take $N=1$,
$L_1=2A+1$, $L_2=0$, and the above quantity reduces to
$$ \phi_Y={(2A+1)!! \over (2A+1)}=(2A-1)!! $$

A further simplification comes from the following remark.
Let $\tilde Y$ denote the Young tableau conjugate to $Y$.
The corresponding conjugate representation of the symmetric
group has same dimension
as the representation encoded in $Y$. It is easy to see that the
conjugate of an even tableau is also even.
Moreover, the Schur function $ch_{\tilde{Y}}(\Gt)$ is obtained
from the Schur function $ch_Y(\Gt)$ by letting the odd $\Gt_{2i+1}$
invariant and changing the even $\Gt_{2i}\to -\Gt_{2i}$.
As a consequence, we have
$\chi_{\tilde{Y}}([2^A])=(-1)^A \chi_Y([2^A])$ and for any even
tableau $Y$
$$ \phi_{\tilde{Y}} = (-1)^{|Y| \over 2} \phi_Y$$
where $\phi_Y$ is defined in lemma 4.

Collecting this information in the generating function $Z$,
with the proviso of lemma 4 and the above remark
at hand in order to simplify
the expression, we get

\bigskip

\noindent{\bf Proposition 1}

$$\encadremath{Z(\Gt,\Gt')=\sum_{A \geq 0} (-1)^{A(A-1) \over 2}
\sum_{Y\equiv l_1>..>l_{2A}
\geq 0 \atop Y\ {\rm even} \ , \ |Y|=2A}
{{\prod_{l \ {\rm odd}} \ l!!\ \prod_{l' \ {\rm even}} (l'-1)!! }
\over {\prod_{l\ {\rm odd,}\ l' \ {\rm even}} (l-l') }}
ch_Y(\Gt) ch_Y(\Gt') }$$

\bigskip

\noindent{\bf 4.} It is to be expected that $Z(\Gt,\Gt')$ can be
expressed as the large $N$ limit of some matrix integral.
We shall write the integral in two forms.
The first one, which we owe to a discussion with I. Kostov,
does not exhibit the symmetry between $\Gt$ and $\Gt'$ (duality
between
vertices and faces). The second one will be
explicitly symmetric.
Let $X$ and $X'$ be diagonal $N \times N$ matrices with non vanishing
diagonal elements (they are therefore invertible) and $M$ a generic
hermitian $N \times N$ matrix, $dM$ the Lebesgue measure on the
$N^2$--dimensional real vector space spanned by $M$. Set
$$\eqalign{
t_k \equiv t_k(X) &= tr(X^k)\ ; \Gt_k\equiv \Gt_k(X)
={tr(X^k)\over k} \cr
t_k' \equiv t_k(X') &= tr(X'^k)\ ; \Gt_k'\equiv \Gt_k(X')
={tr(X'^k)\over k} \cr}$$
Then in the sense of asymptotic (Feynman fatgraphs) expansion in the
variables $\Gt_k(X)$, $\Gt_k(X')$, as $N \to \infty$,  we have

\medskip
\bigskip

\noindent{\bf Proposition 2.}

$$\encadremath{
Z(\Gt(X),\Gt(X'))={{\int dM \exp{-{1 \over 2}tr(MX'^{-1}MX'^{-1})+
\sum_{k=1}^{\infty} \Gt_k(X) tr(M^k)} } \over {\int dM \exp{
-{1 \over 2} tr(MX'^{-1}MX'^{-1})}}}}$$

Each term of the asymptotic expansion is well behaved if for each pair of
diagonal elements $x_i'$, $x'_j$ (where $i$ can be equal to $j$),
we have $Re(x_i' x_j')^{-1}>0$.
This is readily achieved for $X'$ positive definite.
To make the overall integral well defined, we could require that $X$ be
pure imaginary, but this will not cure the divergence of the asymptotic
series (see below).
Since we are only dealing with formal series these subtleties do not
affect the algebraic conclusions.

Consider a graph in the above perturbative expansion.
A vertex of valence $v$ will contribute a factor $t_v(X)=tr(X^v)$.
Each double line (propagator) corresponding to the pair of indices $i$, $j$
is weighted by a factor $x_i'x_j'$.
A "face" of valence $v$ (i.e. bounded
by $v$ edges) and carrying a given index $i$ will be weighted
by a factor $x_i'^v$.
When we sum over all the matrix indices running along the lines of the
fatgraph, a face of valence $v$ contributes a factor
$\sum_i x_i'^v=t_v(X')$.
The remaining symmetry factor will contribute $1/h({\cal G})$,
where $h({\cal G})$ is the order of the automorphism group of
the graph ${\cal G}$. Comparing with the definition \exzz\ of $Z$,
this completes the proof of Proposition 2.

As compared to usual one matrix integrals previously considered in
various models, we note two specificities.

\item{(a)} The "potential" term is completely arbitrary.
This causes no difficulty as we could require that $\Gt_k(X)$ vanish
for large enough $k$.
Indeed at this stage the introduction of the matrix $X$ can be
considered as a trick and we could think of the $\Gt_k$'s as
arbitrary coupling constants. The above interpretation is however
convenient in order to give some flesh to the expression
$ch_Y(\Gt)\equiv ch_Y(\Gt(X))$ in the expansion of proposition 1.

\item{(b)} What is rather unusual is the form of
the quadratic term in the matrix model potential. It involves twice the
matrix $X'$ and is reminiscent of Kontsevich integrals (particularly of
the case $p=3$, corresponding to the Boussinesq hierarchy) although
not exactly the same.
Indeed we are at loss to perform any "angular" average over the unitary
group $U(N)$, which would reduce the $N^2$--dimensional integral to
an $N$--dimensional one over the eigenvalues of the argument $M$.
In this respect it is quite fortunate that the combinatorial treatment
of the fatgraphs enabled us to obtain directly the expansion of $Z$
in the proposition 1 of the previous section.

At this point the reader might wonder: what is the integral representation
of proposition 2 good for?
The answer is twofold.
As we shall see in section 9, it enables one
to obtain an interesting series
for the logarithm of $Z$, the quantity of interest.
On a more low--brow level the integral representation and its
general properties confirm that the sum over {\it connected} fatgraphs
is indeed obtained by taking the logarithm of $Z$
$$ F(\Gt,\Gt')= \log Z(\Gt,\Gt')=\sum_{A \geq 1} F_A(\Gt,\Gt') $$
with
$$ F_A(\Gt,\Gt')=\sum_{\sum v S_v=\sum v F_v=2A} \ t^{\SS} \ t'^{\SF}
\ f(\SS,\SF) $$
where $f$ is defined in \funcf. This could of course be also obtained
through an algebraic treatment by studying the behaviour of $\cal G$
under permutations of identical connected parts.
To calculate $f$'s explicitly, we use the following

\bigskip

\noindent{\bf Lemma 5.}

$$\encadremath{F_A(\Gt,\Gt')
={1 \over A} \sum_{0 \leq p \leq A-1} (-1)^p
\left\vert
\matrix{ Z_{A-p} & Z_{A-p+1} & .&.&.& Z_{A-1}& Z_A \cr
1 & Z_1 & Z_2 & . & . &Z_{p-1}&Z_p \cr
0& 1 & Z_1 & .& .& Z_{p-2} & Z_{p-1} \cr
.&  &  &  &  &  & . \cr
.&  &  &  &  &  & . \cr
.&  &  &  &  &  & . \cr
0& 0 & 0 & . & . & 1& Z_1 \cr }\right\vert}
$$

This performs directly the inversion of Schur polynomials as
$$\exp(\sum_{A \geq 0} u^A F_A) = \sum_{A \geq 0} u^A Z_A
= \sum_{A\geq 0} u^A p_A(F). $$
The proof of the lemma relies on the inverse of Frobenius reciprocity
formula \frob, a direct consequence of the
orthogonality relations for $\GS_{n}$ characters
\eqn\invfrob{ (\Gt_1)^{\nu_1} (2 \Gt_2)^{\nu_2}...(n\Gt_n)^{\nu_n}=
\sum_{Y\ : \ |Y|=\sum_i i \nu_i} \chi_Y(\SN)\  ch_Y(\Gt) }
To use this let us write formally
$F_A=tr(F^A)/A=t_A(F)$,
and express the monomial $At_A(F)=AF_A$ in terms of Schur functions
$\ch_Y(F)$, which, thanks to eqn.\genschur, are just some determinants
of the $p_k(F)=Z_k$. This means that we apply the formula \invfrob\
to the cycle $\SN=[A^1]$, $n=A$.
The corresponding character $\chi_Y([A^1])$ is known to vanish
unless the tableau $Y=Y_{p,q}$,
where $Y_{p,q}$ has one row of say $q+1$ boxes
and $p \geq 0$ rows of one box, $p+q+1=A$ and
$\chi_{Y_{p,q}}([A^1])=(-1)^p$. This leads exactly
to an alternating
sum over determinants with size ranging from $1 \times 1$ to
$A \times A$ and completes the proof of lemma 5.

In spite of signs occuring everywhere, the coefficient
of any monomial $t^{\SS}\ t'^{\SF}$ in $F$
should be a non negative
rational number by construction.
{}From the expression of $Z_A$ obtained in section 3 we can then claim
that we have an "explicit" formula
for the generating function $F$.
However if only in terms of the time (perhaps computer time ...)
it takes
to extract any $f(\SS,\SF)$ from these equations for a graph
with a reasonable size, we realize at once that our task is not complete.
We should rather look for constraints on $F$
which generate faster the required quantities.

A last comment is in order.
{}From the explicit series for $Z(\Gt, \Gt')$ and Cauchy's determinantal
identity, it follows that if $Y$ is an even Young tableau of
size $|Y|=2A$, described as before by a sequence of $l$'s \lis\
$$\eqalign{
\langle ch_Y(M) \rangle &\equiv {{\int dM ch_Y(M)
\exp{-{1 \over 2}tr(MX'^{-1}MX'^{-1})} } \over {\int dM
\exp{-{1 \over 2}tr(MX'^{-1}MX'^{-1})} }} \cr
&=(-1)^{A(A-1) \over 2}
{{\prod_{l \ {\rm odd}} \ l!! \
\prod_{l' {\rm even}} (l'-1)!!} \over {\prod_{l \ {\rm odd} \atop
l' \ {\rm even}} (l-l') }} ch_Y(X') \cr}$$
and the average is zero if $Y$ is not even.
It follows that the Schur functions $ch_Y$ diagonalize the integral
operator, the eigenvalues being zero or the odd integer
$\phi_Y$ computed in lemmas 3-4.
A special case of the above reads
$$ \langle p_{2A}(M) \rangle = (2A-1)!! \ p_{2A}(X') $$
It is obtained by taking a $1 \times 1$ matrix $X=x$, hence
a potential $\sum_{k \geq 1} {x^k \over k}
tr(M^k)$. This also shows the divergent structure of the asymptotic series
when $X$ is one--dimensional.

As suggested above there exists a more symmetric form of the matrix
integral. To obtain it we observe that
$$ \exp{ \sum_{k \geq 1} \Gt_k(X) tr(M^k)}=
\prod_{i=1}^N \exp{\sum_{k \geq 1} x_i^k {tr(M^k) \over k}}
=\prod_{i=1}^N {1 \over \det(1-x_i M)} $$
Parenthetically we see that if all $x_i$ have a bounded positive
imaginary part for instance, this quantity is well defined and is in
fact bounded as $M$ runs over hermitian matrices,
justifying a remark made before.
The inverse determinants can be represented by Gaussian
integral over (complex) $N$--dimensional vectors $v_i$
($i$ indexes the vectors, not their components)
$$ \exp{\sum_{k \geq 1} \Gt_k(X) tr(M^k)} =
{{\int \prod_{i=1}^N dv_i d{\bar {v_i}} \exp{- \sum_{i=1}^N
{\bar {v_i}} (1-x_iM) v_i} } \over { \int \prod_{i=1}^N
dv_i d{\bar {v_i}} \exp{-\sum_{i=1}^N {\bar {v_i}}v_i}}}$$
Inserting this representation into the integral over
$M$ we turn the latter into a Gaussian integral, which is easily
performed.
Indeed defining an $N\times N$ {\it complex} matrix $V$ whose
$i$th column is the vector $v_i$ in components, and integrating
over $M$, the partition function $Z$ takes the form

\bigskip

\noindent{\bf Proposition 3.}

$$\encadremath{
\eqalign{Z(\Gt,\Gt')&=
{{\int dV dV^{\dagger} e^{-tr(VV^{\dagger})+{1 \over 2}tr(VXV^{\dagger}
X')^2} } \over { \int dV dV^{\dagger} e^{-tr(VV^{\dagger})} }} \cr
&={{\int dW dW^{\dagger} e^{-tr(WX^{-1}W^{\dagger}X'^{-1})+
{1 \over 2} tr(WW^{\dagger})^2} } \over
{\int dW dW^{\dagger} e^{-tr(WX^{-1}W^{\dagger}X'^{-1})} } } \cr }} $$

To make sense of these expressions globally we can take here $X$ and $X'$
hermitian positive definite while the coefficient $1 \over 2$ could at
first be replaced by a parameter $\Gl$ purely imaginary or
even better with a negative real part.
In both cases one of the terms in the exponential weight is invariant
under the transformation $V \to g_1 V g_2$ or $W \to g_1 W g_2$,
with $(g_1,g_2) \in U(N)\times U(N)$, revealing the true
symmetry of the problem.
We observe that in the first dissymmetric representation of theorem 2,
we were left with a single $U(N)$ invariance. In other words
we deal with two hermitian matrices $X$ and $X'$, the final result
being conjugation invariant for both.
To integrate over this action on one of them, we had to break the
explicit duality between $X$ and $X'$.

\bigskip

\noindent{\bf 6.} In spite of the pessimistic remarks made above the
formulas obtained so far enable to compute the first few $F_A$'s
in a straightforward way.
On table I, we display a list of the first few $F$'s up to $F_4$
together with the genus of
the corresponding curves in the right column (for each monomial,
it is given by Euler's formula $2-2g=\sum(S_v+F_v) -A$).
To illustrate the use of this table and our criterion of
rationality consider the fatgraph corresponding to the $D_4$ diagram
of genus zero depicted on Fig.1.
It has $3$ vertices of valence $1$, 1 vertex of valence
$3$, $3$ edges and a single face of valence $6$.
It is encoded in $F_3$ in the term
$$ f([1^3 3^1],[6^1])t_1^3 t_3 t_6'=18 f([1^3 3^1],[6^1])\Gt_1^3
\Gt_3 \Gt_6' $$
and we read
$$f([1^33^1],[6^1])={1 \over 3}$$
On the other hand the automorphism group of the graph is
$\IZ / {3 \IZ}$, hence $h([1^3 3^1],[6^1])=3$ and
$$ h([1^3 3^1],[6^1])f([1^3 3^1],[6^1])=1$$
This shows that the corresponding curve is defined over
$\IQ$ as expected.

One can also compute easily a number of other contributions
to $F$. For instance all terms which
only involve $\Gt_1$ and $\Gt_2$ are obtained from a Gaussian integral
and are all of genus zero.
They read
\eqn\trivzer{
\sum_{A \geq 1}  A 2^{A-1} (\Gt_1^2 \Gt_2^{A-1} \Gt_{2A}'+
\Gt_2^A \Gt_A'^2 )}
where the expression has to be symmetrized in $\Gt$, $\Gt'$,
except for the completely symmetric term $\Gt_2^2 \Gt_2'^2$
which appears only once.

Another sequence easily computed corresponds to graphs with only
one face (or only one vertex), which are necessarily connected.
It contributes to $F_A$ through
$$ \Phi_A(\Gt) \Gt_{2A}' + (\Gt \leftrightarrow \Gt') $$
where we omit the symmetrization for the completely symmetric term
$\Gt_{2A} \Gt_{2A}'$ and
$$\eqalign{
\Phi_A(\Gt)&=\sum_{p=0}^{A-1} (-1)^p (2p-1)!! (2A-2p-3)!! \times \cr
&\big[ 2(A-2p-1) \sum_{r=0}^{2p} p_{r}(-\Gt)p_{2A-r}(\Gt)
-(2p+1)p_{2p+1}(-\Gt)p_{2A-2p-1}(\Gt) \big] \cr} $$
{}From this we extract the term, non-zero iff $A$ is even,
of genus $A \over 2$ corresponding to one $2A$--legged
vertex surrounded by  one face of valence $2A$
$$ 2 \sum_{p=0}^{A-1} (-1)^p (2p-1)!! (2A-2p-3)!! (A-2p-1)\Gt_{2A}
\Gt_{2A}' =
{1+(-1)^A \over 2} {2A \over {A+1}} (2A-1)!! \Gt_{2A} \Gt_{2A}'$$
The coefficient is an even integer, in agreement with table I
(we find $4$ for $A=2$, genus $1$ and $168$ for $A=4$, genus $2$).

Pursuing we have a term, non zero iff $A$ is odd, of genus
$A-1 \over 2$ with a vertex of valence one and a
vertex of valence $2A-1$, and only one face of valence $2A$
$$ {1 - (-1)^A \over 2}2(2A-1)!! \Gt_{2A-1} \Gt_1 \Gt_{2A}' \ \ \ {\rm
for} \  A>1$$
For $A=1$ omit the factor $2$, i.e. we get $\Gt_1^2 \Gt_2'$ (as
already noted in $F_1$ on table I).

The next term, non zero for $A$ even, of genus $A-2 \over 2$ reads
$$ {1 +(-1)^A \over 2}2A(2A-3)!! \Gt_1^2\Gt_{2A-2} \Gt_{2A}'\ \ \ {\rm
for} \   \ A>2 $$
and so on...

All the results collected so far suggest the following

\noindent{\bf Conjecture 1.}
$$\encadremath{\vbox{In the expansion of $F$ as an asymptotic
series  of $\Gt$,
$\Gt'$, the coefficient of any monomial is a non-negative
{\it integer}}}$$

Equivalently
$$ \prod_v v^{S_v+F_v} \sum_{{\cal G}_{\SS,\SF}}
{1 \over h({\cal G}_{\SS,\SF})}\ \in \IN.$$

Since as mentioned in section 1, for a connected fatgraph
$h({\cal G}_{\SS,\SF})$ divides $2A$, when it is not zero \BI,
the sum
$$ f(\SS,\SF)=\sum_{{\cal G}_{\SS,\SF}}{1 \over h({\cal G}_{\SS,\SF})}
$$
is a positive rational with denominator a divisor of $2A$.
The conjecture would imply that it can be reduced to the
form
$$ f(\SS,\SF)= { p \over q}$$
where $p$ is an integer and
$q$ is the greatest common divisor of $2A$ and
$\prod_v v^{S_v+F_v}$.
For instance take the term $f([1^1 3^1],[1^1 3^1])t_1t_3t_1't_3'$
with $2A=4$ and $\prod v^{s_v+F_v}=9$ relatively prime.
The conjecture (true in this case) implies that $f$ be an integer
($1$ here).

The above results motivate another conjecture on the form of the
genus $0$ part of $F$, but expanding it on monomials of the form
$${t^{\SS} \over \SS!}{t'^{\SF} \over \SF!}
\equiv \prod_v {t_v^{S_v} \over S_v!}{t'v^{F_v} \over F_v!}$$

\noindent{\bf Conjecture 2.}
$$\encadremath{\vbox{As an asymptotic expansion on monomials
of the form ${t^{\SS}\over \SS!}{t'^{\SF}\over \SF!}$
the genus zero contribution to $F$ has non negative integer
coefficients}}$$

Equivalently
$$ \prod_v S_v!F_v! \sum_{{\cal G}_{\SS,\SF} \atop
{\rm of}\ {\rm genus}\ 0}
{1 \over h({\cal G}_{\SS,\SF})} \  \in \ \IN. $$
We display these integers on table II up to the genus $0$ contribution
to $F_4$.
They are computed from the data of table I.
These integers
are much smaller than those occuring in table I, suggesting that the
second conjecture is sharper than the first one in genus 0.
In higher genus, the above numbers are no longer integers, but
still have relatively small denominators.

\bigskip

\noindent{\bf 7.} The combination of Frobenius formula \frob\
and the expression of characters of the linear group as determinants
of Schur polynomials \genschur\ enables one to perform rather efficiently
some non trivial computations.
Take the example shown on Fig.2 of a graph considered by Osterl\'e
(following Grothendieck,
some authors use the french name "dessin
d'enfant" for what we call here fatgraphs, this applies particularly
well to this example).
This genus $0$ fatgraph corresponds to the
monomial $\Gt_1^4\Gt_3^2\Gt_4\Gt_1'\Gt_{13}'$ in $F_7$, with the same
coefficient in $Z_7$. This is because a graph with two faces, one
of which is of valence $1$, is necessarily connected.
Moreover the only characters $ch_Y(\Gt')$
which have a non vanishing coefficient of $\Gt_1'\Gt_{13}'$ correspond
to Young tableaux with a row of $p$ boxes and $14-p$ rows of one box
or one row with $p\geq 2$ boxes, one row with $2$ boxes and $12-p$ rows
of one box.
In fact the only tableaux actually contributing are $[14^1]$ ($p=14$)
and $[1^{14}]$ ($p=0$) in the first sequence, while only even
$p$'s contribute in the second sequence (they are the only even ones).
Finally we use the (anti)symmetry of the prefactor $\phi_Y$ defined
in lemma 3 under conjugation of the tableau (here $(-1)^A=-1$),
which reduces the computation to four cases, with one
vanishing contribution.
This yields the coefficient of $t_1^4t_3^2t_4t_1't_{13}'$
$$f([1^4 3^2 4^1],[1^1 13^1])={2 \over {3^2.4.13}} \big(
{13!! \over 48}-{13.9!! \over 24}+{13.5!!3!! \over 16} \big)=10.$$
The graph of Fig.2 has an automorphism group reduced to unity.
It follows from section 1 that the number of conjugates of
the corresponding rational curve is at most ten.
As Osterl\'e and collaborators have shown, this number is effectively
$10$ in the present case.

\bigskip

\noindent{\bf 8.} The generating function $Z$ defined in \exzz\
admits a number of specializations. For instance when the matrix
$X'$ is equal to the unit matrix, and $\Gt_k=-x^{k-2}/k$ for $k\geq 3$
while $\Gt_1=\Gt_2=0$, the integral representation of proposition 2
reduces to an integral considered by Penner \PEN\ in the computation
of the virtual Euler characteristic of the mapping class group of
punctured Riemann surfaces (see also \ZI).
More precisely, under the specialization
$$\eqalign{
\Gt_1=\Gt_2=0 &\qquad \Gt_k = -{x^{k-2} \over k} \qquad k \geq 3 \cr
\Gt_k' &= -{y \over k} \qquad k \geq 1 \cr}$$
we obtain for $F=\log Z$ the following series
\eqn\peneq{ F(x,y)= \sum_{n>0,\ 2g-2+n>0} x^{2(2g-2+n)} y^n
{B_{2g} \over {2g(2g-2+n)}} {2g-2+n \choose n} }
with $B_{2g}$ the Bernoulli numbers defined through
$${1 \over 2} + {t \over {e^t -1}} =\sum_{g \geq 0}
B_{2g} {t^{2g} \over (2g)!}.$$
In the sum \peneq\ the integers $n$ and $g$ stand respectively for the
number of faces (there is a factor $y$ for each face) and the genus
of the corresponding fatgraphs. When $g=0$ the coefficient is
understood as a limiting value
$${B_{2g} \over {2g(2g-2+n)}} {2g-2+n \choose n} \bigg\vert_{g=0}
=-{(n-3)! \over n!}. $$
For instance when $n=3$, $g=0$ we get from table I the terms
$$4\Gt_4\Gt_1'^2\Gt_2'+18 \Gt_3^2 \Gt_1'^2\Gt_4'+12\Gt_3^2\Gt_2'^3$$
Under the above specialization this yields
$$4(-{x^2 \over 4})(-y)^2(-{y \over 2})+18(-{x \over 3})^2(-y)^2
(-{y \over 4})+12(-{x \over 3})^2(-{y \over 2})^3
=-{1 \over 6}x^2 y^3$$
in agreement with \peneq.
Similarly for $n=1$, $g=1$, the contributions from table I are
$$ 4 \Gt_4 \Gt_4' +9 \Gt_3^2\Gt_6'$$
which under the specialization yield
$$4(-{x^2 \over 4})(-{y \over 4})+9 (-{x \over 3})(-{y \over 6})
={1 \over 12}x^2 y$$
again in agreement with \peneq\ by using $B_2={1 \over 6}$.

Yet another interesting specialization is when only
finitely many $\Gt$'s are non zero and $X'$ is again the unit matrix.
The integral representation of proposition 2 reduces to the standard
one matrix model which can be dealt with,
using orthogonal polynomial techniques.

\bigskip

\noindent{\bf 9.} Until now no significant use of the integral
representation of proposition 2 was made.
Let us now show how it can be of some help through a saddle
point method.
In general in matrix models the latter is applicable only after
performing the angular integration. The present case is an exception
as we will soon see. Let us look in proposition 2
for stationary points of the integral over $M$ in the numerator.
We find the equation
\eqn\col{ X'^{-1} M X'^{-1}=\sum_{k \geq 1} k \Gt_k(X) M^{k-1}}
For short write $\Gt_k(X)\equiv \Gt_k$.
We are going to solve this equation through the
Lagrange inversion method,
starting from the
term corresponding to $k=1$ on the r.h.s., namely $\Gt_1$id.
The solution $M_0$ thus obtained is readily seen to
commute with $X'$.
Let us recall the Lagrange original problem. For a given analytic
function $\varphi(x,p)$ of $x$, $p$ a parameter,
let us consider the
equation
$$ m=p+\varphi(m,p)$$
and suppose that it has a unique solution $m(p)$ for $p$ small
enough.
Then the function
$$ \partial_x \log[x-p-\varphi(x,p)] $$
is analytic except at $x=m(p)$, where it has a single pole with
residue $1$. We use the Cauchy formula on a contour $C$
surrounding this pole to rewrite
$$\eqalign{ \psi(m(p),p)&= \oint_{C} {dx \over {2i \pi}}
\psi(x,p) \partial_x [ \log(x-p) + \log(1- {\varphi(x,p)
\over (x-p)})] \cr
&=\psi(p,p) + \sum_{k \geq 1} {1 \over {k!}}
\partial_x^{k-1} [\partial_x \psi(x,p)
\varphi(x,p)^k] \big\vert_{x=p} \cr}$$

Returning to our problem, eqn.\col\ is the stationarity condition
for the function
$$\eqalign{ F(M)&=Tr(G(M)) \cr
G(M)&=-{1 \over 2} (MX'^{-1})^2 + \Gt_1 M + V(M) \cr
V(M)&=\sum_{k \geq 2} \Gt_k M^k \cr}$$
Let $P=\Gt_1 X'^2$, hence
$M_0=P(1+{V'(M_0) \over \Gt_1})$, as $M_0$ and $X'$ commute (we can
treat $M_0$ and $P$ as commuting scalars).
Applying the Lagrange method above to the functions
$$\eqalign{
\psi(M,P)=G&=-{1 \over 2} M^2 P^{-1}\Gt_1+M \Gt_1 +V(M) \cr
&={\Gt_1 P \over 2} -{\Gt_1(M-P)^2 P^{-1} \over 2}+V(M) \cr
\varphi(M,P)&={P \over \Gt_1} V'(M), \cr}$$
we finally get
$$\eqalign{
G(M_0,P)&={\Gt_1 P \over 2} +V(P)+\sum_{n=1}^{\infty}
{P^n \over n!\Gt_1^n} \partial_x^{n-1} \bigg[
\big(V'(x)\big)^{n+1}
-{\Gt_1 P^{-1}}(x-P)\big(V'(x)\big)^n \bigg]_{x=P} \cr
&={\Gt_1 P \over 2}+V(P)+\sum_{n=1}^{\infty} {P^n \over
{(n+1)! \Gt_1^n}} \partial_P^{n-1} V'(P)^{n+1}. \cr}$$
To get $F$, we still have to take the trace of the above
expression, i.e. expand it in powers of $P$ and use
$Tr(P^k)=\Gt_1^k Tr(X'^{2k})= 2k\Gt_{2k}' \Gt_1^k$.
This suggests to rewrite the result symbolically,
by setting $P=\Gt_1 \Gt'^2$, acting on $G$ with the
operator $\Gt' \partial_{\Gt'}\equiv 2 P \partial_P$,
and finally substituting $\Gt'^{2k} \to \Gt_{2k}'$.
We have
$$F^{[0]}(\Gt,\Gt')=\Gt_1 P + 2\sum_{n=0}^{\infty}
P\partial_P {P^n \over {(n+1)!\Gt_1^n}} \partial_P^{n-1}
\big(V'(P)\big)^{n+1}\bigg\vert_{P^k=\Gt_1^k \Gt_{2k}'},$$
where the $n=0$ term in the sum is to be understood as
$2 P V'(P)$.
Substituting $V(P)=\sum_{k \geq 2}\Gt_k P^k$, we get
$$\encadremath{
F^{[0]}(\Gt,\Gt')=\sum_{r_1,r_2,.. \geq 0 \atop
2+\sum(i-2)r_i=0} {2 \over \sum r_i} {(\sum r_i)! \over \prod r_i!}
\prod (i\Gt_i)^{r_i} \Gt_{2(r_1+r_2+...-1)}'}$$

A few comments are in order.

\item{(i)} This is the so--called tree
approximation to the logarithm of the integral of proposition 2.
A tree here means a fatgraph of genus zero with only one face
(only one $\Gt'$ appears in each monomial of the
expansion of $F^{[0]}$, and the genus is given by the Euler
formula $2-2g=1+\sum r_i-(\sum r_i -1)=2$, with $S=\sum r_i$
vertices, $F=1$ face and $A=\sum r_i -1=\sum i r_i /2$
edges).
Therefore $F^{[0]}$ is only the tree piece of the genus
zero energy $F^{(0)}$ of table II.

\item{(ii)} The coefficients appearing in $F^{[0]}$ are
integers generalizing the Catalan numbers, in agreement with
conjecture 1. To prove it,
write in various ways
$$ {2 \over \sum r_i} { (\sum r_i)! \over \prod r_i!}=
{2 \over r_k} {(\sum r_i -1)! \over
{(r_k-1)! \prod_{i \neq k} r_i!}} $$
therefore for all non vanishing $r_k$'s,
\eqn\catal{
{2 \over \sum r_i} \times ({\rm integer})={2 \over r_k}
\times ({\rm integer})}
and the $r$'s are subject to the constraint
$2+ \sum (i-2)r_i=0$. The l.h.s. of \catal\ is a rational
with a denominator dividing all $r_k$'s, hence a divisor of $2$
due to the constraint. If one of the $r_k$'s at least is odd,
the denominator is one, and the number \catal\ is an integer.
If all the $r_k$'s are even, the denominator of \catal\ has
to divide all $r_k/2$, hence is one, again due to the constraint.
In all cases we proved that \catal\ is an integer.
It is a generalization of the Catalan numbers ${2n \choose n}/(n+1)$.
In terms of these numbers the connected function for genus zero
tree--fatgraphs defined in \funcf\
reads
$$f([{\smash \prod} i^{r_i}],[(2({\smash \sum} r_i -1))^1])={1 \over 2A}
{2 \over \sum r_i} {(\sum r_i)! \over \prod r_i!}$$
with the constraint $2+\sum (i-2)r_i =0$, and
$2A=\sum ir_i=2(\sum r_i -1)$.

\item{(iii)} The expression of $F^{[0]}$ agrees also with
conjecture 2. To see why, notice that $\sum r_i \geq 2$
and rewrite
$$ {2 \over \sum r_i} {(\sum r_i)! \over \prod r_i!}
\bigg(\prod(i \Gt_i)^{r_i}\bigg) \Gt_{2(\sum r_i -1)}'=
(\sum r_i -2)! \bigg(\prod {t_i^{r_i} \over r_i!}\bigg)
t_{2(\sum r_i -1)}'.$$

\item{(iv)} The expression of $F^{[0]}$ simplifies in the
case of tree fatgraphs with only one-- and $k$--valent
vertices
$$\eqalign{F^{[0]}(\Gt_1,0,..,0,\Gt_k,0,..;\Gt')&=\cr
&\sum_{n \geq 0}
{2 \over {2+n(k-1)}} {2+n(k-1) \choose n} \Gt_1^{2+n(k-2)}
(k \Gt_k)^n \Gt_{2+2n(k-1)}' \cr}$$
In the particular case of the star fatgraph ($A$ vertices of valence
$1$ and $1$ vertex of valence $A$, corresponding to the term
$n=1$ in the above sum) we find
$$f([1^A A^1],[2A^1]) = {1 \over A}$$
which exhibits the cyclic group $\IZ_{A}$ as automorphism group.

So far we only performed the tree approximation to the integral of
proposition 2. Let us now proceed and compute the
"one loop" corrections
by the semi--classical approximation at the above
stationary point $M_0$.
We have to study
$$ Z= { \int dM e^{S(M,X')} \over {\int dM e^{S_0(M,X')} }}$$
with
$$\eqalign{
S(M,X')&=-{1 \over 2}Tr(MX'^{-1})^2 + \sum_{k \geq 2} \Gt_k Tr(M^k)\cr
S_0(M,X')&=-{1 \over 2}Tr(MX'^{-1})^2. \cr}$$
Semi--classically, one expands $S$ up to second order
around the stationary point $M_0$
$$ S(M,X')=F^{[0]} + {1 \over 2}
\sum_{\Ga,\Gb,\Gc,\Gd} (M-M_0)_{\Ga,\Gb}
{\partial^2 S \over{\partial M_{\Ga,\Gb} \partial M_{\Gc,\Gd}}}(M_0)
(M-M_0)_{\Gc,\Gd} + ...$$
and integrates over the Gaussian corrections.
Taking into account the contribution from the denominator, we get
$$\eqalign{ \log Z &= F^{[0]} +F^{[1]}+... \cr
F^{[1]}&= -{1 \over 2} \log \det (Q R^{-1}), \cr }$$
where $Q$ and $R$ are linear operators acting on the space of hermitian
matrices, defined by
$$\eqalign{ Q \ M &= X'^{-1} M X'^{-1} -\sum_{k \geq 2} k \Gt_k
\sum_{r,s \geq 0 \atop r+s=k-2 } M_0^r M  M_0^s \cr
R \ M &= X'^{-1} M X'^{-1} \cr
R^{-1} \ M &= X' M X'. \cr}$$
Set $QR^{-1} = 1 - K$, where
\eqn\defopk{
K \  M = \sum_{k \geq 2} t_k \sum_{r,s \geq 0 \atop r+s=k-2}
M_0^r X' M  X' M_0^s.  }
{}From now on we return to the notation $t_k$ for $k\Gt_k=Tr(X^k)$.
The one loop correction reads
$$ F^{[1]}= -{1 \over 2} \log \det (1-K) = {1 \over 2} \sum_{N \geq 1}
{Tr(K^N) \over N}. $$
Using the definition of the operator $K$ and the fact that $M_0$ and $X'$
commute, we get
\eqn\fonel{\eqalign{
F^{[1]}&= {1 \over 2} \sum_{N \geq 1} { 1 \over N}
\sum_{r_1,..,r_N,s_1,..,s_N \geq 0}
\big(\prod_{i=1}^N t_{r_i+s_i+2}\big)\times \cr
&\times Tr(X'^N M_0^{r_1+..+r_N}) Tr(X'^N M_0^{s_1+..+s_N}). }}
To compute $M_0^r$, we use the Lagrange method again, with the functions
$\psi(M,P)=M^r$ and
$\varphi(M,P)=(P/ t_1)V'(M)=P\sum_{k \geq 2}
(t_k/t_1) M^{k-1}$, $P=t_1 X'^2$, and we get
$$\eqalign{
M_0^r &= P^r + \sum_{n=1}^{\infty} {P^n \over n! t_1^n} \partial_P^{n-1}
\big[ r P^{r-1} V'(P)^n \big] \cr
&=t_1^r X'^r +r \sum_{n=1}^{\infty} \sum_{k_1,..,k_n \geq 1}
t_1^{k_1+..+k_n+r-n} t_{k_1+1} ..t_{k_n+1} \times \cr
&\times {X'^{2(k_1+..+k_n+r)} \over {(k_1+..+k_n+r)}}
{k_1+..+k_n+r \choose n}. \cr}$$
Let us reorganize the summation over $k_i$ into a sum over the
$n_j=|\{i, \ k_i=j \}|$ for $j \geq 2$ and $n_1=\sum k_i -n+r$,
$n= \sum n_i$, $\sum k_i=\sum i n_i$,
and take the trace after multiplication by $X'^N$
$$Tr(M_0^r X'^N)=r \sum_{n_1,n_2,..\geq 0 \atop
r+\sum (i-2) n_i =0}
{(\sum n_i -1)! \over { \prod n_i!}}
\big( \prod t_i^{n_i} \big)
t_{N+2 \sum r_i}'$$
where we understand the sum as yielding $t_N'$ when $r=0$ (this can
be summarized in the convention $0\times (-1)!=1$).
Substituting this into \fonel, we have
$$\encadremath{\eqalign{
F^{[1]}&= \sum_{N \geq 1} {1 \over 2N}
\sum_{r_a, \ s_a \geq 0,\atop a=1,..,N}
(\sum r_a)(\sum s_a)t_{r_1+s_1+2}...t_{r_N+s_N+2} \times \cr
&\times\sum_{n_i \geq 0 \atop \sum r_a +\sum (i-2)n_i=0}
\sum_{m_i \geq 0 \atop \sum s_a +\sum (i-2)m_i=0}
{(\sum n_i -1)!\over{ \prod n_i!}}
{(\sum m_i -1)! \over{ \prod m_i!}}  \times \cr
&\times t_1^{n_1+m_1} t_2^{n_2+m_2}t_3^{n_3+m_3}...
t_{N+2\sum n_i}'
t_{N+2\sum m_i}'\cr}}$$
Here we use again the convention $0 \times (-1)!=1$ when
$\sum n_i=0$ (resp. $\sum m_i=0$), which implies $\sum r_a=0$
(resp. $\sum s_a=0$).

Of course the contributions in $F^{[0]}+F^{[1]}$ up to $A=4$
agree with the data in tables I and II.
Remarkably the expression involves only genus zero fatgraphs!
This is readily seen by applying Euler's formula with
$F=2$ faces (two $t'$), $S=\sum(n_i+m_i)+N$
vertices, and $A=[\sum (r_a+s_a)+\sum i( n_i+m_i)+2N]/2=S$
edges, hence $2-2g=F-A+S=2$.
We do not fully understand this phenomenon.
Perhaps higher genus contributions would correspond to
other non trivial saddle points.
We could pursue the semi--classical expansion beyond
"one loop" order. But the expressions become quite cumbersome.

\bigskip

\noindent{\bf 10. } In this note we gathered information
about the sum of the inverse orders of
the automorphism groups of fatgraphs with specified valences
of faces and vertices. We gave a compact expression for the
generating function for general fatgraphs.
The connected case however proved to be more subtle.
Although in principle we just had to take the logarithm $F$ of
the previous generating function, we were not able
to find such a compact and ready--to--use expression in that
case.

The introduction of the matrix integral of proposition 2 should
shed some light on the problem of calculating directly the
generating function $F$ for connected fatgraphs.  We were only able
to perform a semi--classical expansion around a small stationary
point of the matrix model action, yielding apparently only genus
zero contributions to $F$ (this remains for us slightly mysterious).
But one should try to extract more information from this
matrix model, presumably by implementing the equations of motion
in an efficient way.

\vfill\eject
\centerline{\bf Table I}

$$\eqalignno{
\ &\    &{\rm genus}  \cr
F_1 &= \Gt_1^2 \Gt_2'+ \Gt_2 \Gt_1'^2     &0 \cr
F_2&= 4 \Gt_4 \Gt_4' &1 \cr
&+4\Gt_1^2\Gt_2 \Gt_4'+ 4 \Gt_4 \Gt_1'^2 \Gt_2'
+4 \Gt_2^2\Gt_2'^2+9\Gt_1\Gt_3\Gt_1'\Gt_3'  &0 \cr
F_3&= 30 \Gt_1 \Gt_5 \Gt_6' + 30 \Gt_6 \Gt_1' \Gt_5'
+24 \Gt_2 \Gt_4 \Gt_6'+24 \Gt_6\Gt_2'\Gt_4'
+9 \Gt_3^2 \Gt_6'+9 \Gt_6 \Gt_3'^2  &1 \cr
&+12 \Gt_1^2 \Gt_2^2 \Gt_6'+12 \Gt_6 \Gt_1'^2 \Gt_2'^2
+20 \Gt_1^2 \Gt_4 \Gt_1' \Gt_5'+20 \Gt_1 \Gt_5 \Gt_1'^2 \Gt_4'
+30 \Gt_1 \Gt_2 \Gt_3 \Gt_1' \Gt_5'&0 \cr
&+ 30 \Gt_1 \Gt_5 \Gt_1' \Gt_2'\Gt_3'
+48 \Gt_1 \Gt_2 \Gt_3 \Gt_2' \Gt_4'+48 \Gt_2 \Gt_4
\Gt_1' \Gt_2' \Gt_3'
+18 \Gt_3^2 \Gt_1'^2 \Gt_4'+18 \Gt_1^2 \Gt_4 \Gt_3'^2
&0 \cr
&+12 \Gt_2^3 \Gt_3'^2+12 \Gt_3^2 \Gt_2'^3
+6 \Gt_1^3 \Gt_3 \Gt_6'+6 \Gt_6 \Gt_1'^3\Gt_3' &0 \cr
F_4&= 168 \Gt_8 \Gt_8' &2 \cr
&+120 \Gt_1^2 \Gt_6 \Gt_8'+120 \Gt_8 \Gt_1'^2 \Gt_6'+
240 \Gt_1\Gt_2\Gt_5 \Gt_8'+240 \Gt_8 \Gt_1' \Gt_2' \Gt_5'
+245 \Gt_1 \Gt_7 \Gt_1' \Gt_7' &1 \cr
&+192 \Gt_1 \Gt_3 \Gt_4  \Gt_8'+192 \Gt_8 \Gt_1' \Gt_3' \Gt_4'
+96 \Gt_2^2 \Gt_4 \Gt_8'+96 \Gt_8 \Gt_2'^2 \Gt_4'
+72 \Gt_2 \Gt_3^2 \Gt_8'+72 \Gt_8 \Gt_2' \Gt_3'^2 &1 \cr
&+168 \Gt_2 \Gt_6 \Gt_1' \Gt_7'+168 \Gt_1 \Gt_7 \Gt_2' \Gt_6'
+210 \Gt_3 \Gt_5 \Gt_1' \Gt_7'+210 \Gt_1 \Gt_7 \Gt_3' \Gt_5'
+216 \Gt_2 \Gt_6 \Gt_2' \Gt_6' &1 \cr
&+112 \Gt_4^2 \Gt_1' \Gt_7' +112 \Gt_1 \Gt_7 \Gt_4'^2
+180 \Gt_3 \Gt_5 \Gt_2' \Gt_6'+180 \Gt_2 \Gt_6 \Gt_3' \Gt_5'
&1 \cr
&+96 \Gt_4^2 \Gt_2'\Gt_6'+96 \Gt_2 \Gt_6 \Gt_4'^2
+96 \Gt_4^2 \Gt_4'^2+225 \Gt_3 \Gt_5 \Gt_3' \Gt_5' &1 \cr
&+48 \Gt_1^3 \Gt_2 \Gt_3 \Gt_8'+48 \Gt_8 \Gt_1'^3 \Gt_2' \Gt_3'
+32 \Gt_1^2 \Gt_2^3 \Gt_8'+32 \Gt_8 \Gt_1'^2 \Gt_2'^3
+54 \Gt_1^2 \Gt_6 \Gt_1'^2 \Gt_6' &0 \cr
&+35\Gt_1^3 \Gt_5 \Gt_1' \Gt_7'+35 \Gt_1 \Gt_7 \Gt_1'^3 \Gt_5'
+112 \Gt_1^2 \Gt_2\Gt_4 \Gt_1' \Gt_7'+112 \Gt_1 \Gt_7 \Gt_1'^2
\Gt_2' \Gt_4'+ 162 \Gt_2 \Gt_3^2\Gt_2'\Gt_3'^2  &0 \cr
&+63 \Gt_1^2 \Gt_3^2 \Gt_1' \Gt_7'+63 \Gt_1 \Gt_7 \Gt_1'^2 \Gt_3'^2
+84 \Gt_1 \Gt_2^2 \Gt_3 \Gt_1' \Gt_7'
+84\Gt_1 \Gt_7\Gt_1'\Gt_2'^2\Gt_3'
+300 \Gt_1 \Gt_2 \Gt_5 \Gt_1' \Gt_2' \Gt_5' &0 \cr
&+96 \Gt_1^2 \Gt_2 \Gt_4 \Gt_2' \Gt_6'
+ 96\Gt_2 \Gt_6 \Gt_1'^2 \Gt_2' +54 \Gt_1^2 \Gt_3^2 \Gt_2' \Gt_6'
+54 \Gt_2 \Gt_6 \Gt_1'^2 \Gt_3'^2
+32 \Gt_2^4 \Gt_4'^2
+32 \Gt_4^2 \Gt_2'^4 &\ \ \ 0 \cr
&+144 \Gt_1 \Gt_2^2 \Gt_3 \Gt_2' \Gt_6'
+144\Gt_2\Gt_6\Gt_1'\Gt_2'^2\Gt_3'
+144 \Gt_1 \Gt_3 \Gt_4\Gt_1'^2 \Gt_6'+144\Gt_1^2\Gt_6\Gt_1'\Gt_3'\Gt_4'
&0 \cr
&+60 \Gt_1 \Gt_2 \Gt_5 \Gt_1'^2\Gt_6'+60\Gt_1^2\Gt_6\Gt_1'\Gt_2'\Gt_5'
+54 \Gt_2 \Gt_3^2 \Gt_1'^2 \Gt_6'+54\Gt_1^2\Gt_6\Gt_2'\Gt_3'^2
+288\Gt_1\Gt_3\Gt_4\Gt_1'\Gt_3'\Gt_4'\ \ \ \ \ \ \ &\ \ \ 0 \cr
&+75\Gt_1^3\Gt_5\Gt_3'\Gt_5'+75\Gt_3\Gt_5\Gt_1'^3\Gt_5'
+120\Gt_1^2\Gt_2\Gt_4\Gt_3'\Gt_5'+120\Gt_3\Gt_5\Gt_1'^2\Gt_2'\Gt_4'
+128 \Gt_2^2 \Gt_4 \Gt_2'^2\Gt_4'  &0 \cr
&+180 \Gt_1 \Gt_2^2 \Gt_3\Gt_3'\Gt_5'
+180\Gt_3\Gt_5\Gt_1'\Gt_2'^2\Gt_3' +180 \Gt_2 \Gt_3^2\Gt_1'\Gt_2'\Gt_5'
+180\Gt_1\Gt_2\Gt_5\Gt_2'\Gt_3'^2 &0 \cr
&+120 \Gt_1 \Gt_3\Gt_4\Gt_1'\Gt_2'\Gt_5'
+120\Gt_1\Gt_2\Gt_5\Gt_1'\Gt_3'\Gt_4'+192\Gt_1 \Gt_3\Gt_4\Gt_2'^2\Gt_4'
+192\Gt_2^2\Gt_4\Gt_1'\Gt_3'\Gt_4' &0 \cr
&+64 \Gt_1^2\Gt_2\Gt_4\Gt_4'^2+64\Gt_4^2\Gt_1'^2\Gt_2'\Gt_4'
+72\Gt_1^2\Gt_3^2\Gt_4'^2+72\Gt_4^2\Gt_1'^2\Gt_3'^2
+8\Gt_1^4\Gt_4\Gt_8'+8\Gt_8\Gt_1'^4\Gt_4' &0 \cr}$$

\vfill\eject

\centerline{\bf Table II}

$$\eqalign{
F_1^{(0)}&= {t_1^2 \over 2!} t_2'+ t_2 {t_1'^2 \over 2!} \cr
F_2^{(0)}&={t_1^2 \over 2!}t_2 t_4'+ t_4 {t_1'^2 \over 2!}t_2'
+{t_2^2 \over 2!}{t_2'^2 \over 2!}+t_1 t_3 t_1' t_3' \cr
F_3^{(0)}&=t_1 t_2 t_3 t_1' t_5'+t_1 t_5 t_1' t_2' t_3'
+t_1 t_2 t_3 t_2' t_4'+t_2 t_4 t_1' t_2' t_3'
+2{t_1^3 \over 3!} t_3 t_6' \cr
&+2t_6 {t_1'^3 \over 3!}t_3'
+2{t_1^2 \over 2!}{t_2^2 \over 2!}t_6'
+2 t_6 {t_1'^2 \over 2!}{t_2'^2 \over 2!}
+2 {t_1^2 \over 2!}t_4 t_1' t_5'+2t_1 t_5{t_1'^2 \over 2!}t_4'\cr
&+2 {t_3^2 \over 2!}{t_1'^2 \over 2!}t_4'
+2 t_4 {t_1^2 \over 2!} {t_3'^2 \over 2!}
+2 {t_2^3 \over 3!}{t_3'^2 \over 2!}
+2 {t_3^2 \over 2!}{t_2'^3 \over 3!}\cr
F_4^{(0)}&={t_1^2 \over 2!}t_2 t_4 t_1' t_7'
+ t_1 t_7{t_1'^2 \over 2!}t_2' t_4'+t_1 t_3 t_4 t_1' t_2' t_5'
+t_1 t_2 t_5 t_1' t_3't_4'+t_1 t_3 t_4 {t_1'^2 \over 2!}t_6'\cr
&+{t_1^2 \over 2!}t_6 t_1' t_3' t_4'
+2t_1 {t_2^2 \over 2!}t_3 t_1' t_7'
+2t_1 t_7 t_1' {t_2'^2 \over 2!} t_3'
+2{t_1^2 \over 2!}t_2 t_4 t_2' t_6'
+2 t_2 t_6 {t_1'^2 \over 2!}t_2' t_4'\cr
&+2 {t_1^2 \over 2!}{t_3^2 \over 2!}t_2' t_6'
+2 t_2 t_6 {t_1'^2 \over 2!}{t_3'^2 \over 2!}
+2t_1 {t_2^2 \over 2!}t_3 t_2' t_6'
+2 t_2 t_6 t_1' {t_2'^2 \over 2!}t_3'
+2 t_1 t_2 t_5 {t_1'^2 \over 2!} t_6'\cr
&+2{t_1^2 \over 2!} t_6 t_1' t_2' t_5'
+2{t_1^2 \over 2!}t_2 t_4 t_3' t_5'
+2 t_3 t_5 {t_1'^2 \over 2!}t_2' t_4'
+2 t_1 {t_2^2 \over 2!} t_3 t_3' t_5'
+2t_3 t_5 t_1' {t_2'^2 \over 2!} t_3'\cr
&+2 t_2 {t_3^2 \over 2!} t_1' t_2' t_5'
+2t_1 t_2 t_5  t_2' {t_3'^2 \over 2!}
+2 {t_1^2 \over 2!} t_2 t_4 {t_4'^2 \over 2!}
+2 {t_4^2 \over 2!} {t_1'^2 \over 2!} t_2' t_4'
+2 t_1 t_3 t_4 t_1' t_3' t_4'\cr
&+2t_2 {t_3^2 \over 2!}{t_1'^2 \over 2!}t_6'
+2{t_1^2 \over 2!}t_6 t_2' {t_3'^2 \over 2!}
+2 {t_2^2 \over 2!} t_4 t_1' t_3' t_4'
+2 t_1 t_3 t_4  {t_2'^2 \over 2!} t_4'
+2 {t_2^2 \over 2!}t_4 {t_2'^2 \over 2!}t_4'\cr
&+2 t_2 {t_3^2 \over 2!}t_2' {t_3'^2 \over 2!}
+3t_1 t_2 t_5 t_1' t_2' t_5'
+4{t_1^2 \over 2!}{t_3^2 \over 2!} t_1' t_7'
+4 t_1 t_7{t_1'^2 \over 2!}{t_3'^2 \over 2!}
+4 {t_1^2 \over 2!}{t_3^2 \over 2!}{t_4'^2 \over 2!}\cr
&+4 {t_4^2 \over 2!}{t_1'^2 \over 2!}{t_3'^2 \over 2!}
+6{t_1^4 \over 4!}t_4 t_8'+6t_8 {t_1'^4 \over 4!}t_4'
+6{t_1^3 \over 3!}t_2 t_3 t_8'
+6t_8 {t_1'^3 \over 3!}t_2' t_3'\cr
&+6{t_1^2 \over 2!}{t_2^3 \over 3!} t_8'
+6 t_8 {t_1'^2 \over 2!}{t_2'^3 \over 3!}
+6 {t_1^3 \over 3!} t_5 t_1' t_7'
+6 t_1 t_7  {t_1'^3 \over 3!} t_5'
+6 {t_1^2 \over 2!}t_6 {t_1'^2 \over 2!}t_6'\cr
&+6 {t_1^3 \over 3!} t_5 t_3' t_5'
+6 t_3 t_5 {t_1'^3 \over 3!} t_5'
+6 {t_2^4 \over 4!}{t_4'^2 \over 2!}
+6 {t_4^2 \over 2!} {t_2'^4 \over 4!} \cr
}$$
\vfill\eject

\hbox{}

\vskip 6.0in

\centerline{\bf Figure Captions}

\noindent{\bf Fig.1:} The genus 0 graph "$D_4$", with $A=3$,
$\SS=[1^3 3^1]$ and $\SF=[6^1]$.

\noindent{\bf Fig.2:} The genus 0 "dessin d'enfant", with $A=7$,
$\SS=[1^4 3^2 4^1]$ and $\SF=[1^1 13^1]$.

\listrefs
\end